\documentclass[aps,prb,twocolumn,noeprint]{revtex4-1}

\usepackage[bookmarks=false]{hyperref}
\hypersetup{colorlinks=true, citecolor=blue, urlcolor=blue, linkcolor=blue}
\usepackage{graphicx}
\usepackage{epsfig}
\usepackage{amsmath}
\usepackage{graphics}
\usepackage{epsfig}
\usepackage{amsmath}
\usepackage{amsfonts}
\usepackage{amssymb}
\usepackage{xcolor,cancel}
\usepackage{bm}
\usepackage{float}
\usepackage{subfiles}

\begin{document}


\title{ Timescale separation solution of  Kadanoff-Baym equations for  quantum transport in time-dependent fields}

\author{Thomas D. Honeychurch} 
\author{Daniel S. Kosov}
\affiliation{College of Science and Engineering, James Cook University, Townsville, QLD, 4811, Australia}

\begin{abstract}

The interaction with time-dependent external fields, especially the interplay between time-dependent driving and quantum correlations, changes the familiar picture of electron transport through nanoscale systems.
Although the exact solution of the problem of AC quantum transport of noninteracting electrons has been known for more than two decades, the  treatment of correlated particles presents a significant theoretical challenge.
In this paper, using the perturbative separation of fast electron tunnelling and slow driving  time-scales, we  developed a practical approach for time-dependent quantum transport with nonequilibrium Green's functions. The fast electronic dynamics is associated with relative time whilst the slow driving is related to the central time in the  Green's functions. The ratio of characteristic electron tunneling time over the period of harmonic driving is used  as a small parameter in the theory to obtain  a convergent time-derivative expansions of the Green's functions. This enables the algebraic solution of  the Kadanoff-Baym equations in Wigner space. Consequently, we produced analytical expressions for dynamical corrections to advanced, retarded, and lesser Green's functions, as well as an improved expression for AC electric current.
The method developed is applicable to the general case of multi-channel electron transport  through a correlated central region.  The theory is applied to different transport scenarios: time-dependent transport through a driven single-resonant level is compared to exact results; and electron transport through a molecular junction described by the Holstein model with a time-oscillating voltage bias is also investigated.
\end{abstract}

\maketitle

\section{Introduction}

A nanoscale electronic  junction is typically a single molecule attached to two macroscopic leads. In the standard scenario, the static voltage bias is applied to the lead electrodes and the electric current across the molecule is measured. Whilst interest in time-dependent quantum transport phenomena has a long history \cite{Platero2004}, recently, microwave or optical irradiation of the molecular scale electronic junctions have been used to control electronic properties of the system\cite{doi:10.1063/1.2159491,doi:10.1021/nl062273j,PhysRevLett.97.067006}. 
 Access to the optical range is of particular importance, since the frequencies of the time-dependent driving become comparable to molecular vibrational frequencies, which suggests the opportunity 
to stir the electron-vibrational dynamics in molecular junctions.
These recent experimental advances have resulted in an increase in the ongoing development and application of time-dependent quantum transport theories tailored specifically to the properties of molecular electronic junctions \cite{Galperin2012,PhysRevB.96.085425,Peskin2017,Ridley2017,Ochoa2015}.

Historically, Tien and Gordon proposed their seminal qualitative theory to describe the effects of harmonic time-dependent potential on electronic tunnel junctions as early as 1963\cite{Tien1963}. {     Since this work}, further more sophisticated approaches, based on Floquet  and  scattering
theories \cite{Pedersen1998,Grifoni1998,hanggi2003,Kohler2005,blanter2007,moskalets2002,moskalets2004,moskalets-book,moscalets2014},  quantum master equations\cite{PhysRevB.96.085425,Peskin2017}, and notably  nonequilibrium Green's functions (NEGF) based approaches\cite{Arrachea2005,Arrachea2006,Ke2010,Tuovinen2014,Wang1999,Sun1998,Jauho1994,Datta1992,Stafford1996,Brandes1997,Stefanucci2004,Pastawski1992,Chen1991,Maciejko2006,Zhu2005,Ridley2015,Ridley2017,nazarov2007,vonOppen2016} have been developed to deal with time dependent transport in driven quantum systems. The use of NEGF {     allows for the possibility of a systematic treatment of correlations in time-dependent electron transport. In particular, the effects of electron-phonon interactions with time-dependent junctions have been investigated\cite{Souto2018,Ueda2017,Avriller2019,Ding2014,Zhang2015,Ding2016,Souto2015}.} However, this often comes with a considerable increase in complexity, computationally and theoretically, with the helpful simplifications of the noninteracting case being inapplicable.

In this paper, we develop an approach which algebraically solves the Kadanoff-Baym equations for a molecular electronic junction in time-dependent fields using the separation of time-scales between fast electronic tunelling and slow oscillations of driving external fields. It enables us to produce analytic expressions for dynamical corrections to adiabatic (instantaneous) Green's functions and, consequently, dynamical corrections to adiabatic expressions for observable quantities such as time-dependent electric current. The approach follows the ideas of  Arrachea and Moskalets in that it makes use of adiabatic approximations to solve the NEGF Dysons equations \cite{Arrachea2006}. Brandes  based his truncation method for Green's functions in driven junctions \cite{Brandes1997} on the same physical arguments, albeit a technically different realization; he uses discrete Fourier transform with respect to the central time in Green's functions which is restricted to small frequencies -- that is qualitatively equivalent to our central time derivatives gradient expansion of Wigner space Kadanoff-Baym equations.

The paper is organized as follows. Section II describes how the gradient expansion is used to separate time-scales in the Kadanoff-Baym equations, converting the differential equations to a set of algebraic equations which can be resolved analytically. In section II, we also produce analytical expressions for the dynamical corrections to all Green's functions and the electric current. Section  III contains  applications of the proposed theory to time-dependent transport through a harmonically driven single-resonant level and through a molecular junction described by the Holstein model with a time-oscillating voltage bias.  
Section IV summarizes the main results of the paper.

Natural units for quantum transport  are used throughout the paper,  with $\hbar$, $e$ and $k_{B}$ set to unity.

\section{Theory}

\subsection{Hamiltonian, Green's functions and self-energies}

In this section, we present the Hamiltonian of the system, give the basic definitions of Green's functions and self-energies, and introduce notations that will be used throughout the paper. We start with the tunnelling Hamiltonian for quantum transport through an open quantum system; it consists of a central region connected to two macroscopic leads. {    Within this work, the central region is a molecule as realised in molecular electronics}, however all our results are applicable to the general case where  it is represented by a quantum dot, atom, or  any other nano-scale quantum system.
The Hamiltonian is
\begin{equation}
H(t)=H_M(t) + H_L(t) +H_R(t) +H_{ML}(t) + H_{MR}(t),
\label{hamiltonian}
\end{equation}
where $H_M$ is the Hamiltonian for the molecule, $H_L$  and  $H_R$ are the Hamiltonians for the left and  right leads, while  $H_{ML}$  and $H_{MR}$ are for the interaction between the central region and the left and right leads, respectively. 
All terms in the Hamiltonian may be time-dependent and we  do not make any assumptions about the form of this time-dependence at this stage.

The molecular Hamiltonian  is taken in the most general form as
\begin{equation}
H_M= \sum_{ij} ({     t_{ij}(t)} + V_{ij}(t)) d^\dag_i d_{j} +v_c=  \sum_{ij} h_{ij}(t)d^\dag_i d_{j}  + v_c,
\end{equation}
 The  quantities {    $t_{ij}(t)$}  and $V_{ij}(t) $ are matrix elements of the single-particle part of the molecular Hamiltonian and the external time-dependent driving potential, respectively. The electron-electron interaction  and electron-phonon interaction are designated collectively by $v_c$. The particular form of $v_c$ is not relevant for our formal derivations and will be presented later, when we apply the theory to specific examples. Creation $d^\dag_i$ and annihilation  $d_i$ operators are for an electron in the molecular single-particle state $i$.
 
The left and right leads  are modelled as macroscopic reservoirs of noninteracting electrons
\begin{equation}
\label{leads}
H_L +H_R =  \sum_{ k\alpha=L,R}  \epsilon_{k\alpha}(t)  c^{\dagger}_{k\alpha} c_{k\alpha},
\end{equation}
where $a^{\dagger}_{k\alpha}$($a_{k\alpha}$) creates (annihilates) an electron in the single-particle state $k$ of either the left ($\alpha=L$) or  the right ($\alpha=R$) lead. The coupling between central region and left and right leads         {is} given by the tunnelling interaction
\begin{equation}
\label{coupling}
H_{ML}+H_{MR}=  \sum_{i k \alpha=L,R } (t_{k\alpha i}(t) c^{\dagger}_{k \alpha} d_i +\mbox{h.c.} ),
\end{equation}
where $t_{k\alpha i}(t)$ is the time-dependent tunnelling amplitudes between  leads and molecular single-particle states.

 In the NEGF formalism, the terms which we wish to evaluate are given by the full retarded, advanced and lesser Green's functions, which are defined in standard way as\cite{haug2008}:
\begin{equation}
{\cal G}_{ij}^R(t,t') = -i \theta(t-t') \langle \{d_i (t), d^\dag_j (t')\} \rangle,
\end{equation}
\begin{equation}
{\cal G}_{ij}^A(t,t') = [ {\cal G}_{ij}^R(t',t) ]^*
\end{equation}
and
\begin{equation}
{\cal G}_{ij}^<(t,t') = i \langle d^\dag_j (t')d_i (t) \rangle.
\end{equation}

The self-energies due to lead-molecule coupling are
\begin{equation}
\label{eq:lead_self-energy}
\Sigma_{\alpha ij}^{R,A,<}(t,t') = \sum_{k} t^*_{k\alpha  i}(t) g_{k \alpha}^{R,A,< } (t,t')    t_{k \alpha j }(t'),
\end{equation}
where the Green's functions for the separated leads with time-dependent single-particle energy levels are given by the following expressions\cite{haug2008}
\begin{equation} 
\label{g<}
g_{k \alpha}^{<}(t,t')= i f_{k \alpha}e^{-i \int_t^{t'}\epsilon_{k\alpha}(t_1) dt_1},
\end{equation}
\begin{equation} 
\label{gA}
g_{k \alpha}^{A}(t,t')=i\theta(t'-t) e^{-i \int_t^{t'}\epsilon_{k\alpha}(t_1) dt_1}, 
\end{equation}
{    
\begin{equation} 
\label{gR}
 g_{k \alpha}^{R}(t,t') =  g_{k \alpha}^{A}(t',t)^*.
\end{equation}}
The calculations of Fermi-Dirac occupation numbers $f_{k \alpha }$ in the lesser Green's functions of the leads deserves special discussion. To develop a NEGF transport theory, we make a standard assumption that in the infinite past the system and the leads are separated from each other and then the interaction between them is turned on. Moreover, we also assume that the system is described by the static Hamiltonian in the infinite past, before the time-dependent perturbation starts to act on the system, followed by the coupling of the regions. Therefore, 
 the Fermi-Dirac occupation numbers in the free lesser Green's function of the leads  (\ref{g<})  should be computed using the  static parts of single-particle energies 
 \begin{equation}
f_{k\alpha}= \frac{1}{1+e^{(\epsilon_{k\alpha} -\mu_\alpha)/T_\alpha}},
\end{equation}
where $\mu_\alpha$ and $T_\alpha$ are the chemical potential and the temperature for lead $\alpha$.

\subsection{Gradient expansion of Wigner space Kadanoff-Baym equations}

We begin with the Kadanoff-Baym equations of motion for the retarded, advanced and lesser Green's functions \cite{haug2008}.
The Kadanoff-Baym equations are
\begin{multline}
\Big( i \partial_t - \mathbf h(t) \Big) \bm{\mathcal G}^{<}(t,t') = \int dt_1 \Big[ \mathbf {\Sigma}_{\text{tot}}^R(t,t_1)  \bm{\mathcal G}^<(t_1,t') 
\\
+ \mathbf \Sigma_{\text{tot}}^{<}(t,t_{1}) \bm{\mathcal G}^A(t_1,t') \Big]
 \label{eq:KD<}
\end{multline}
and 
\begin{multline}
\Big( i  \partial_t  - \mathbf h(t) \Big) \bm{\mathcal G}^{R/A}(t,t') = 
 \mathbf I \delta(t-t') \\
 +  \int dt_1  \mathbf \Sigma_{\text{tot}}^{R/A}(t,t_{1})  \bm{\mathcal G }^{R/A}(t_{1},t').
  \label{eq:KDA/R}
\end{multline}
We have chosen to work with the Kadanoff-Baym equations in matrix form where the Green's functions $\bm{\mathcal G }$, self-energies $\mathbf {\Sigma}$ and Hamiltonian $\mathbf h$ are matrices in the molecular single-particle space. Here $\mathbf I$ is the identity matrix in molecular space. We introduced the total self-energy
\begin{equation}
\mathbf {\Sigma}^{R/A/<}_{\text{tot}} = \mathbf {\Sigma}^{R/A/<} +  \mathbf {\Sigma}^{R/A/<}_c,
\end{equation}
where $\mathbf \Sigma_c$ is the self-energy from correlations in the central region and the choice of the particular form is not relevant for our immediate discussion in this section.

We define central time $T$ and relative  time $\tau$ for the Green's function ${\cal G }(t,t')$ as
\begin{equation}
T = \frac{1}{2}(t+t'), \;\;\;
\tau = t - t',
\end{equation}
and introduce the Wigner representation of   two-time functions  in the Kadanoff-Baym equations
\begin{equation}
{\widetilde A}(\omega,T) = \int^{+\infty}_{-\infty}  d\tau \; e^{i \omega \tau} A (t,t')
\end{equation}
{     and}
\begin{equation}
A(t,t')= \frac{1}{2 \pi} \int^{+\infty}_{-\infty}  d\omega \; e^{-i \omega \tau}  \widetilde A(\omega,T).
\end{equation}

Upon the application of the Wigner transformation to the convolution $A(t,t')=\int dt_{1}B(t,t_{1})C(t_{1},t')$, we find that 
\begin{equation}
\widetilde A(T,\omega)=e^{-\frac{i}{2}\left(\partial_{T}^{B}\partial_{\omega}^{C}-\partial_{\omega}^{B}\partial_{T}^{C}\right)} \widetilde B(T,\omega) \widetilde C(T,\omega). 
\end{equation}
Here and throughout the rest of the paper, the partial differentials (i.e. $\partial^A_b$) are with respect to $b$ acting on $A$
and the tilde symbol denoting that the function is in Wigner space.

The fast and slow timescales in a driven molecular junction are easily detectable in the Wigner representation.
The external driving is associated with explicit central time dependence, whereas electronic dynamics is linked with relative time variations. This means that 
for all Green's functions $\bm{\mathcal G }(t,t')$, the slow time-dependent external field implies that $\bm{\mathcal G }(T+\tau/2,T-\tau/2)$ varies slowly with the central time $T$, but oscillates fast with the relative time $\tau$. 
This idea has been often used to separate classical (slow) and quantum (fast) degrees of freedom using NEGF theory\cite{Bode12,catalysis12,kershaw18,kershaw17,kershaw19,subotnik17-prl,Yadalam2016,Stefanucci2018,galperin2019}.

Transforming the Kadanoff-Baym equations (\ref{eq:KD<},\ref{eq:KDA/R}) to the Wigner representation removes all time-integrals via an infinite series of time and energy derivatives collected in the exponential operators:
\begin{multline}
\left[\frac{i}{2}  \partial_T +\omega-e^{-\frac{i}{2}\partial^{\mathcal G}_{\omega}\partial^h_{T}}\bm h \right]  \widetilde{\bm{\mathcal G}}^{<}(T,\omega) 
=
\\
e^{-\frac{i}{2}\left(\partial_{T}^{\Sigma}\partial_{\omega}^{\mathcal G}-\partial_{\omega}^{\Sigma}\partial_{T}^{\mathcal G}\right)} 
\Big( 
\widetilde{\bm{\Sigma}}_{\text{tot}} ^{<}(T,\omega) \widetilde{\bm{\mathcal G}}^{A}(T,\omega) 
\\
+\widetilde{\bm{\Sigma}}_{\text{tot}}^{R}(T,\omega)  \widetilde{\bm{\mathcal G}}^{<}(T,\omega) 
\Big)
\label{eq:KD<_wigner}
\end{multline}
and
\begin{multline} 
\left[\frac{i}{2} \partial_T +\omega-e^{-\frac{i}{2}\partial^{\mathcal G}_{\omega}\partial^h_{T}}\bm h \right] \widetilde{\bm{\mathcal G}}^{A/R} (T,\omega) 
= 
\mathbf I 
\\
+ e^{-\frac{i}{2}\left(\partial_{T}^{\Sigma}\partial_{\omega}^{\mathcal G}-\partial_{\omega}^{\Sigma}\partial_{T}^{ \mathcal G}\right)} \widetilde{\bm{\Sigma}}_{\text{tot}}^{A/R} (T,\omega) \widetilde{\bm{\mathcal G}}^{A/R} (T,\omega).
\label{eq:KDA/R_wigner}
\end{multline}
The equations of motion (\ref{eq:KD<_wigner}, \ref{eq:KDA/R_wigner}) are governed  by explicitly time-dependent Hamiltonian and describe  the exact  evolution of the lesser, advanced and retarded Green's functions. 
The direct numerical solution of Kadanoff-Baym equation with explicitly time-dependent Hamiltonian represents a formidable computational challenge and possible only for very simplified systems\cite{Stefanucci2013}.
Our goal is to develop an approximate theory employing the separation of time-scales and assuming that the rate of change of the system Hamiltonian due to external driving is smaller than the electronic tunneling time across the junction. That means that the small parameter  is 
\begin{equation}
\Omega / \Gamma \ll 1,
\end{equation}
 where $\Omega$ is a characteristic external  driving frequency and $\Gamma$ is the molecular level broadening due to the coupling to the leads.

To solve Kadanoff-Baym equations approximately,  we expand the exponent, keeping the first order terms in central time derivatives:

\begin{widetext}
\begin{equation} \label{eq:KD<_wigner_first_1}
\left[\frac{i}{2}\partial_T+\omega-\bm h \right] \widetilde{\bm{\mathcal G}}^{<} 
+\frac{i}{2}\partial_T \bm h \partial_\omega \widetilde{\bm{\mathcal G}}^{<} 
= \widetilde{\bm{\Sigma}}^{R}_{\text{tot}} \widetilde{\bm{\mathcal G}}^{<} 
+\widetilde{\bm{\Sigma}}^{<}_{\text{tot}} \widetilde{\bm{\mathcal G}}^{A} 
-\frac{i}{2} \left(\partial_{T}^{\Sigma}\partial_{\omega}^{{\mathcal G}}
-\partial_{\omega}^{\Sigma}\partial_{T}^{{\mathcal G}}\right) \widetilde{\bm{\Sigma}}^{R}_{\text{tot}} \widetilde{\bm{\mathcal G}}^{<} 
-\frac{i}{2} \left(\partial_{T}^{\Sigma}\partial_{\omega}^{{\mathcal G}}-\partial_{\omega}^{\Sigma}\partial_{T}^{{\mathcal G}}\right) \widetilde{\bm{\Sigma}}^{<}_{\text{tot}} \widetilde{\bm{\mathcal G}}^{A} 
\end{equation}

\begin{equation}
\left[\frac{i}{2}\partial_T+\omega-\bm{h}  \right] \widetilde{\bm{\mathcal G}}^{A/R}
+\frac{i}{2}\partial_T \bm h \partial_\omega \widetilde{\bm{\mathcal G}}^{A/R}
= \bm I 
+ \widetilde{\bm{\Sigma}}^{A/R}_{\text{tot}} \widetilde{\bm{\mathcal G}}^{A/R} 
- \frac{i}{2} \left(\partial_{T}^{\Sigma}\partial_{\omega}^{\mathcal G}
-\partial_{\omega}^{\Sigma}\partial_{T}^{\mathcal G}\right)\widetilde{\bm{\Sigma}}_{\text{tot}}^{A/R} \widetilde{\bm{\mathcal G}}^{A/R} 
 \label{eq:KDA/R_wigner_first_1}
\end{equation}
\end{widetext}

{     For brevity, we omit the functional dependence on $T$ and $\omega$ in the Green's functions, Hamiltonian matrix and self-energies.}
{      The noninteracting Green's functions for the leads (\ref{g<},\ref{gA}, \ref{gR}), which enter into the corresponding self-energies,  are computed exactly.}

 All Wigner space molecular Green's functions are expanded with respect its change with time
\begin{equation}
\label{expansionG}
 \widetilde{\bm{\mathcal G}}= \widetilde{\bm{\mathcal G}}_{(0)}+\widetilde{\bm{\mathcal G}}_{(1)}+O( (\Omega/\Gamma)^2), 
\end{equation}
such that $\widetilde{\bm{\mathcal G}}_{(1)}$ is linear in $\partial_{T}$, and the remaining terms involves the higher order derivatives with respect to central time $T$. Here, we truncate the expansion to the first order.
Considering the correlation self-energy as a functional of Green's function, we expand it up to the first order as well
\begin{equation}
\label{expansionSigma}
 \widetilde{\bm{ \Sigma}}_{c}[ \widetilde{\bm{\mathcal G}}]\approx  \underbrace{ \widetilde{\bm{\Sigma}}_{c}[\widetilde{\bm{\mathcal G}}_{(0)}]}_{ \widetilde{\bm{\Sigma}}_{c(0)}}+
 \underbrace{
 \Big(\widetilde{\bm{ \Sigma}}_{c}[\widetilde{\bm{\mathcal G}}_{(0)}+\widetilde{\bm{\mathcal G}}_{(1)}] - \widetilde{\bm{\Sigma}}_{c}[\widetilde{\bm{\mathcal G}}_{(0)}]] \Big).
 }_{ \widetilde{\bm{\Sigma}}_{c(1)}}
\end{equation}

Making  expansions (\ref{expansionG}) and (\ref{expansionSigma})  for  Green's functions and self-energies in  Eq. (\ref{eq:KDA/R_wigner_first_1}) and
collecting the zeroth  order terms, we get 
\begin{equation} \label{eq:GA/R_0}
\widetilde{\bm{\mathcal G}}^{A/R}_{(0)} =\left[ \omega-\bm h-\bm{\widetilde{\Sigma}}^{A/R} - \bm{\widetilde{\Sigma}}^{A/R}_{c(0)} \right]^{-1}
\end{equation}
Therefore, the zeroth order term is the standard Green's function computed adiabatically for time-dependent molecular Hamiltonian $\mathbf h(T)$ and time-dependent leads self-energies $ \bm{\widetilde{\Sigma}}^{A/R}_{\text{tot}}(T,\omega)$.
Notice that the zeroth order correction is not static, it changes in adiabatically in time, instantaneously following the external time-dependent perturbation.

Similarly, we collect the first order terms to obtain the dynamical corrections for advanced/retarded Green's functions:
\begin{equation} \label{eq:GA/R_1}
\begin{split}
\widetilde{\bm{\mathcal G}}^{A/R}_{(1)} 
=\widetilde{\bm{\mathcal G}}^{A/R}_{(0)} \bm{\widetilde{\Sigma}}^{A/R}_{c(1)}  \widetilde{\bm{\mathcal G}}^{A/R}_{(0)}  
\\
-\frac{i}{2} \widetilde{\bm{\mathcal G}}^{A/R}_{(0)} \partial_T  \widetilde{\bm{\mathcal G}}^{A/R}_{(0)} 
-\frac{i}{2} \widetilde{\bm{\mathcal G}}^{A/R}_{(0)}\partial_T \bm h  \partial_\omega \widetilde{\bm{\mathcal G}}^{A/R}_{(0)} 
\\
-\frac{i}{2} \widetilde{\bm{\mathcal G}}^{A/R}_{(0)}\partial_T \left( \bm{\widetilde{\Sigma}}^{A/R} + \bm{\widetilde{\Sigma}}^{A/R}_{c(0)} \right) \partial_\omega \widetilde{\bm{\mathcal G}}^{A/R}_{(0)}
\\
+\frac{i}{2} \widetilde{\bm{\mathcal G}}^{A/R}_{(0)}\partial_\omega \left( \bm{\widetilde{\Sigma}}^{A/R} + \bm{\widetilde{\Sigma}}^{A/R}_{c(0)} \right) \partial_T  \widetilde{\bm{\mathcal G}}^{A/R}_{(0)}
\end{split}
\end{equation}
for which we computed explicitly the central-time derivatives, 
\begin{equation} 
\partial_{T} \widetilde{\bm{\mathcal G}}^{A/R}_{(0)} 
=  \widetilde{\bm{\mathcal G}}^{A/R}_{(0)} \partial_T \left(\bm h + \bm{\widetilde{\Sigma}}^{A/R} + \bm{\widetilde{\Sigma}}^{A/R}_{c(0)} \right) \widetilde{\bm{\mathcal G}}^{A/R}_{(0)},
\label{eq:dtG_0^A/R}
\end{equation}
and also the derivatives with respect to energy
\begin{equation}
\partial_{\omega} \widetilde{\bm{\mathcal G}}^{A/R}_{(0)} 
=-\left(\widetilde{\bm{\mathcal G}}^{A/R}_{(0)} \right)^{2}
+ \widetilde{\bm{\mathcal G}}^{A/R}_{(0)} \partial_\omega \left( \bm{\widetilde{\Sigma}}^{A/R} + \bm{\widetilde{\Sigma}}^{A/R}_{c(0)} \right) \widetilde{\bm{\mathcal G}}^{A/R}_{(0)}.
 \label{eq:domegaG_0^A/R}
\end{equation}

Similarly, using the expansions (\ref{expansionG}) and (\ref{expansionSigma})   in the truncated Kadanoff-Baym equation (\ref{eq:KD<_wigner_first_1}), we arrive at the zeroth order lesser Green's function
\begin{equation}  \label{eq:G^<_{(0)}}
\widetilde{\bm{\mathcal G}}^{<}_{(0)}
=\widetilde{\bm{\mathcal G}}^{R}_{(0)}\bm{\widetilde{\Sigma}}^{<}\widetilde{\bm{\mathcal G}}^{A}_{(0)} + \widetilde{\bm{\mathcal G}}^{R}_{(0)}\bm{\widetilde{\Sigma}}^{<}_{c(0)}\widetilde{\bm{\mathcal G}}^{A}_{(0)}.
\end{equation}
The zeroth order corrections constitute the adiabatic solution to the problem, which is equivalent to imbuing the static solution with  time-dependent parameters.

For the first order terms, we find, 
\begin{widetext}
\begin{equation}\label{eq:G^<_{(1)}}
\begin{split}
\widetilde{\bm{\mathcal G}}^{<}_{(1)}
 = 
-\frac{i}{2} \widetilde{\bm{\mathcal G}}^{R}_{(0)} \partial_T \widetilde{\bm{\mathcal G}}^{<}_{(0)}
-\frac{i}{2} \widetilde{\bm{\mathcal G}}^{R}_{(0)} \partial_T \bm h \partial_\omega  \widetilde{\bm{\mathcal G}}^{<}_{(0)}
+\widetilde{\bm{\mathcal G}}^{R}_{(0)} \left(\bm{\widetilde{\Sigma}}^{<} + \bm{\widetilde{\Sigma}}^{<}_{c(0)}\right) \widetilde{\bm{\mathcal G}}^{A}_{(1)}
\\
-\frac{i}{2} \widetilde{\bm{\mathcal G}}^{R}_{(0)} \partial_T  \left( \bm{\widetilde{\Sigma}}^{<} + \bm{\widetilde{\Sigma}}^{<}_{c(0)}\right) \partial_\omega\widetilde{\bm{\mathcal G}}^{A}_{(0)}
+\frac{i}{2} \widetilde{\bm{\mathcal G}}^{R}_{(0)} \partial_\omega \left(\bm{\widetilde{\Sigma}}^{<} +  \bm{\widetilde{\Sigma}}^{<}_{c(0)}\right) \partial_T \widetilde{\bm{\mathcal G}}^{A}_{(0)}
\\
-\frac{i}{2} \widetilde{\bm{\mathcal G}}^{R}_{(0)} \partial_{T} \left( \bm{\widetilde{\Sigma}}^{R} +  \bm{\widetilde{\Sigma}}^{R}_{c(0)} \right) \partial_{\omega}\widetilde{\bm{\mathcal G}}^{<}_{(0)}
+\frac{i}{2} \widetilde{\bm{\mathcal G}}^{R}_{(0)} \partial_{\omega} \left( \bm{\widetilde{\Sigma}}^{R} +  \bm{\widetilde{\Sigma}}^{R}_{c(0)} \right)  \partial_{T}\widetilde{\bm{\mathcal G}}^{<}_{(0)}.
\end{split}
\end{equation}
\end{widetext}
The Eqs. (\ref{eq:GA/R_0}),(\ref{eq:GA/R_1}),(\ref{eq:G^<_{(0)}}) and (\ref{eq:G^<_{(1)}}) represent       solutions for the Kadanoff-Baym equations under the assumption that the characteristic period of the external time-dependent driving is slower than the time spent by the electron in the central region.

\subsection{Dynamical  corrections to time-dependent electric current} 

Having obtained the non-adiabatic corrections to the retarded, advanced  and lesser Green's         {function,} we are now ready to obtain the time-dependent electric current. Let us begin with the general expression for electric current flowing into the molecule from $\alpha=L,R$  lead at time $t$\cite{haug2008}:
\begin{multline}
{J}_{\alpha}(t) =2 \text{Re} \int  dt_1
\text{Tr} \Big\{{ \bm{\mathcal G}}^<(t,t_1)\mathbf \Sigma_{\alpha}^A(t_1,t)
\\
+{\bm{\mathcal G}}^R(t,t_1)  \mathbf \Sigma_{\alpha}^<(t_1,t) \Big\}.
\label{mw}
\end{multline}
Here the trace is taken over the molecular single-particle states. Transforming this equation to the Wigner space, we find  

\begin{equation}
\begin{split}
J_{\alpha}(t) 
=  2 \operatorname{Re} \int^{\infty}_{-\infty} \frac{d\omega}{2\pi}\operatorname{Tr}   [\mathbf C_\alpha(t,\omega)] ,
\end{split}
\end{equation}
where $\mathbf C_\alpha(t,\omega) $ is defined as
\begin{multline}
\mathbf C_\alpha(t,\omega)= e^{-\frac{i}{2} \left(\partial^G_t \partial^\Sigma_\omega - \partial^G_\omega \partial^\Sigma_t  \right)} 
\Big( \widetilde{\bm{\mathcal G}}^<(t,\omega) \widetilde{\bm{\Sigma}}^A_\alpha (t,\omega) 
\\
+ \widetilde{\bm{\mathcal G}}^R(t,\omega) \widetilde{\bm{\Sigma}}^<_\alpha(t,\omega) \Big).
\end{multline}

As before,  we truncate the exponential terms, assuming slow variation with respect to central time of the Green's functions and self-energies in Wigner representation:
\begin{equation}
\begin{split}
\bm C_\alpha 
= \widetilde{\bm{\mathcal G}}^< \widetilde{\bm{\Sigma}}^A_\alpha 
+ \widetilde{\bm{\mathcal G}}^R \widetilde{\bm{\Sigma}}^<_\alpha
-\frac{i}{2}\partial_t \widetilde{\bm{\mathcal G}}^< \partial_\omega \widetilde{\bm{\Sigma}}^A_\alpha  
\\
+\frac{i}{2}\partial_\omega\widetilde{\bm{\mathcal G}}^< \partial_T \widetilde{\bm{\Sigma}}^A_\alpha
-\frac{i}{2}\partial_t \widetilde{\bm{\mathcal G}}^R \partial_\omega \widetilde{\bm{\Sigma}}^<_\alpha
+\frac{i}{2}\partial_\omega \widetilde{\bm{\mathcal G}}^R \partial_T \widetilde{\bm{\Sigma}}^<_\alpha.
\end{split}
\end{equation}

Splitting the zeroth and first order contributions, we calculate the adiabatic current  and first order dynamical corrections to it. For the zeroth order,  the adiabatic current, we find that 
\begin{equation} \label{eq:current_zeroth}
J_{L,(0)}(t)= 2  \operatorname{Re}  \int^{\infty}_{-\infty} \frac{d\omega}{2\pi} \operatorname{Tr} \big[ \widetilde{\bm{\mathcal G}}^<_{(0)}\widetilde{\bm{\Sigma}}^A_L
+  \widetilde{\bm{\mathcal G}}^R_{(0)} \widetilde{\bm{\Sigma}}^<_L  \big].
\end{equation}
The first order correction to the electric current is
\begin{equation} \label{eq:current_first}
\begin{split}
J_{\alpha,(1)}(t)
= 2  \operatorname{Re}  \int^{\infty}_{-\infty} \frac{d\omega}{2\pi} \operatorname{Tr}  \big[ \widetilde{\bm{\mathcal G}}^<_{(1)} \widetilde{\bm{\Sigma}}^A_\alpha 
+\widetilde{\bm{\mathcal G}}^R_{(1)} \widetilde{\bm{\Sigma}}^<_\alpha 
\\ 
-\frac{i}{2}\partial_T \widetilde{\bm{\mathcal G}}^<_{(0)} \partial_\omega \widetilde{\bm{\Sigma}}^A_\alpha
+\frac{i}{2}\partial_\omega \widetilde{\bm{\mathcal G}}^<_{(0)} \partial_T \widetilde{\bm{\Sigma}}^A_\alpha)
\\
-\frac{i}{2}\partial_T \widetilde{\bm{\mathcal G}}^R_{(0)} \partial_\omega \widetilde{\bm{\Sigma}}^<_\alpha
+\frac{i}{2}\partial_\omega \widetilde{\bm{\mathcal G}}^R_{(0)} \partial_T \widetilde{\bm{\Sigma}}^<_\alpha
\big].
\end{split}
\end{equation}

If the wide-band approximation is assumed (see appendix A for details on the use of wide-band approximation with time-dependent leads chemical potentials and tunneling coupling), the first order correction to the current reduces to the following: 
\begin{equation} \label{eq:current_first}
\begin{split}
J_{\alpha,(1)}(t)= 
2 
\operatorname{Re} 
\int^{\infty}_{-\infty} \frac{d\omega}{2\pi} \operatorname{Tr} \big[\widetilde{\bm{\mathcal G}}^<_{(1)} \widetilde{\bm{\Sigma}}^A_\alpha 
+\widetilde{\bm{\mathcal G}}^R_{(1)} \widetilde{\bm{\Sigma}}^<_\alpha
\\
-\frac{i}{2}\partial_T \widetilde{\bm{\mathcal G}}^R_{(0)} \partial_\omega \widetilde{\bm{\Sigma}}^<_\alpha
+\frac{i}{2}\partial_\omega \widetilde{\bm{\mathcal G}}^R_{(0)} \partial_T \widetilde{\bm{\Sigma}}^<_\alpha
\big].
\end{split}
\end{equation}

\section{Results}

\subsection{AC current through single resonant-level}
Let us first consider the simple case of a single resonant-level connected to two leads with harmonically  driven chemical potentials. 
In this case the molecular Hamiltonian is 
\begin{equation}
H_M=(\epsilon +V(t)) d^\dagger d,
\end{equation}
where $\epsilon$ is the energy of the resonant-level, $V(t)$ is the external time-depending potential, and $d^\dagger (d) $  are creation(annihilation) operators for a resonant-level electron.
The altering of chemical potentials is equivalent to uniform harmonic modulations of the leads' single-particle energies
\begin{equation}
\label{harmonic}
\epsilon_{k\alpha}(t) = \epsilon_{k\alpha} + \Delta_\alpha \cos \Omega t ,
\end{equation}
where $ \Delta_\alpha$ is the amplitude of the         {oscillations of the energy levels and}  $\Omega$ is the modulation frequency. We assume 
that the energy of resonant-level is also shifted in time with the same frequency due to the time-oscillating voltage drop between         {the} leads or due to a time-dependent gate 
\begin{equation}
V(t) =  \Delta \cos \Omega t.
\end{equation}

\begin{figure}
\includegraphics[width=1.0\columnwidth]{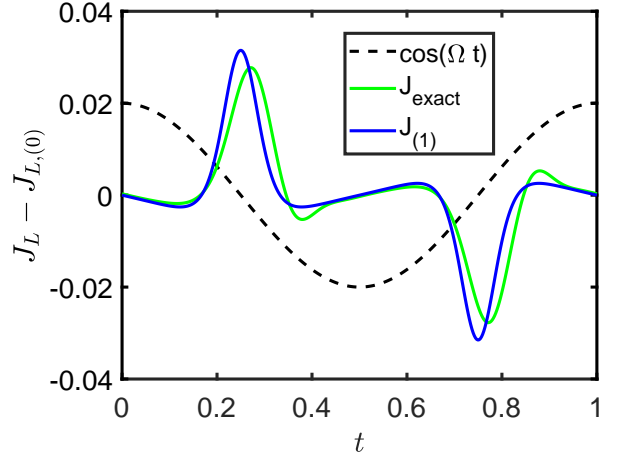}
\caption{Comparison between the first order correction to current and the difference between the exact  and zeroth order currents. The parameters are  $\Gamma_L = \Gamma_R =  0.5\Gamma$, $\Delta_L  = \Gamma$, $\Delta_R = -\Gamma$, $\Delta = 0$, $\epsilon = 0$, $\Omega = 0.1\Gamma $, $\mu_L = \mu_R = 0$ and $T = 0.001\Gamma$. 
The current is measured in units of $\Gamma$ and time is given in periods of the external driving $2 \pi/\Omega$.}
  \label{fig:plot1_1_2}
\end{figure}

We employ the wide-band approximation for         {the} leads,         {which} brings the self-energies to         {a} particularly simple forms (appendix A).
Under         {this approximation,} the exact solution for the model is known\cite{Jauho1994} and the proposed theory will be benchmarked against         {this} available exact         {result}.

{     We assess {the} adiabatic, zeroth order, electric current  $J_{(0)}(t)$  given by Eq.(\ref{eq:current_zeroth}) along with the first order dynamical correction $J_{(1)}(t)$ given by Eq.(\ref{eq:current_first}) against  the exact time-dependent current  $J_{\text{exact}}(t)$ from literature\cite{Jauho1994}.}  It is difficult to visually judge the performance of the method from the comparison of the total currents, since it is dominated by the adiabatic contribution in both cases. In order to eliminate this bulk trivial         {term,} we subtract $J_{(0)}$ from  $J_{\text{exact}}(t)$ and plot it along with our first order dynamical correction
$J_{(1)}(t)$ -- the result is shown in figure  \ref{fig:plot1_1_2}.  One can see that the first order correction captures the essential dynamical features of the exact solution
{     which is associated with the retardation of the filling and emptying of the resonant level when
it crosses the Fermi level of leads}.

\begin{figure}
\includegraphics[width=1.0\columnwidth]{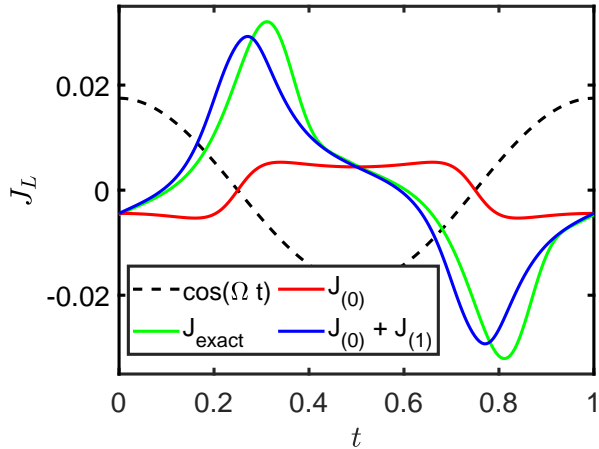}
\caption{Current through a single-resonant level as a function of time. The parameters are  $\Gamma_L = \Gamma_R =  0.5\Gamma$, $\Delta_L  = 0$, $\Delta_R = 0.125 \Gamma$, $\Delta = 1$, $\epsilon = 0$, $\Omega = 0.1\Gamma $, $\mu_L = \mu_R = 0$ and $T = 0.1\Gamma$. The current is measured in units of $\Gamma$ and time is given in periods of the external driving $2 \pi/\Omega$.}
  \label{fig:plot2_3}
\end{figure}

Figure \ref{fig:plot2_3}  shows that in certain situations  the first order dynamical correction plays a critical role. Let us consider the case when the left lead is grounded at all time, a small AC voltage is applied to the right lead whilst a much larger harmonically oscillating gate voltage is applied to the central region. The adiabatic current does not show any dynamical characters and fails to reproduce the exact time behavior; it simply follows instantaneously the applied time dependent voltage bias. The first order dynamical correction significantly improves the results by bringing the value of         {the} current close to the exact value.

\subsection{Holstein model in time-dependent Hartree approximation}

Having checked the proposed theoretical approach by the comparison with exact results, we turn our attention to electron transport through the molecular junction with electron-vibration interaction. The molecule  is described by the Holstein model Hamiltonian:
\begin{equation}
H_{M}(t)= \epsilon  d^\dag d  + \lambda (a^\dag + a) d^\dag d +  \omega a^\dag a,
\end{equation}
where $\epsilon$ molecular orbital energy, $\omega$ is molecular         {vibrational} energy, and $\lambda$ is the strength of the electron-vibration coupling. The operator $d^\dagger (d) $  creates (annihilates) an electron on molecular orbital, and         {$a^\dagger (a)$} is         {the} bosonic creation (annihilation) operator for the molecular vibrations. The electronic spin does not play any physical role in this section and will not be included explicitly into the equations.
We note that the molecule is still coupled to two leads with Hamiltonians in Eq.(\ref{leads}), lead-molecular interactions  in Eq.(\ref{coupling}), and         {the} total system Hamiltonian in Eq.(\ref{hamiltonian}). Similar to the single resonant-level considered in section III-A, the         {single particle energies of the leads} depend on time harmonically as given in Eq.(\ref{harmonic}).

Assuming that the vibration is slower than the characteristic electron tunnelling time $\omega \ll \Gamma$,         {meaning} that at any given time moment a tunnelling electron interacts with a constant static molecular vibration, the correlation self-energy can be written in Hartree approximation as \cite{ryndyk-book,galperin05}
\begin{equation}
\Sigma^{A/R}_{c} (t,t') = \chi n(t) \delta \left(t-t' \right) ,
\end{equation}
\begin{equation}
\Sigma^{<}_{c} (t,t') = 0 ,
\end{equation}
where 
\begin{equation}
\chi = - \frac{2\lambda^2}{\omega} 
\end{equation}
and $n(t)$ is time-dependent molecular population is given by
\begin{equation}
{     n(t) = \operatorname{Im} \left[ \int^{\infty}_{-\infty} \frac{d\omega}{2\pi} \widetilde{{\mathcal G}}^< (t,\omega) \right] .} 
\label{eq:occupation_corrections}
\end{equation}
{     The assumption that $ \omega \ll \Gamma$ does not effect or related with the gradient expansion, as unlike $\Omega$, the phonon frequency  $\omega$  does not dictate the time-dependent variation of the system. It is required for the validity of  the Hartree approximation.}

The corresponding correlation self-energies in the Wigner space representation         {become}
\begin{equation}
{ \widetilde  \Sigma}^{A/R}_{c} (T,\omega) = \chi n(T),
\end{equation}
\begin{equation}
\widetilde{\Sigma}^{<}_{c}  (T,\omega) = 0.
\end{equation}

\begin{figure}
\includegraphics[width=1.0\columnwidth]{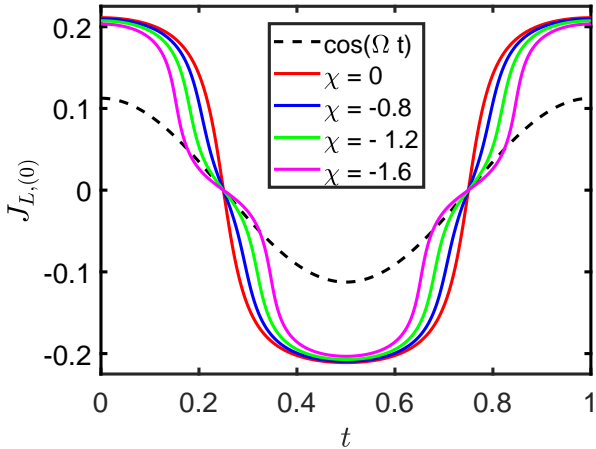}
\caption{The adiabatic time-dependent current computed with varying values for $\chi$.  The other parameters are  $\Gamma_L = \Gamma_R =  0.5\Gamma$, $\Delta_L  = 2 \Gamma$, $\Delta_R = -2 \Gamma$, $\epsilon= 0$, $\Omega = 0.05\Gamma $, $\mu_L = \mu_R = 0$ and $T = 0.001\Gamma$.
The current is measured in units of $\Gamma$ and time is given in periods of the external driving $2 \pi/\Omega$. }
    \label{fig:plotMF5}
\end{figure}

\begin{figure}
\includegraphics[width=1.0\columnwidth]{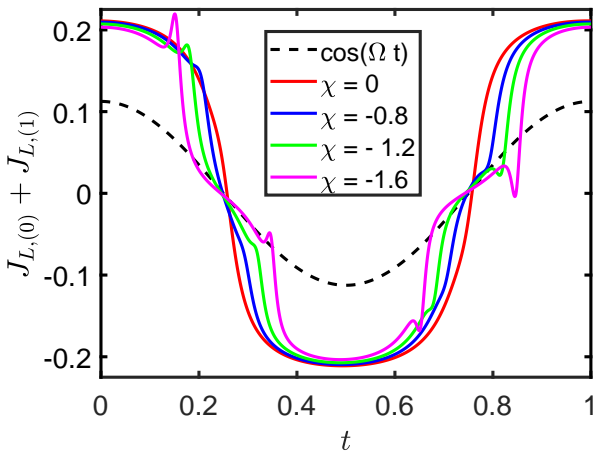}
\caption{Time-dependent current with the first order dynamical corrections  computed with varying values for $\chi$. The parameters are the same as in Fig.  (\ref{fig:plotMF5}). }
    \label{fig:plotMF6}
\end{figure}

In order to compute the self-energies, and consequently the Green's functions, we need to know the adiabatic molecular electronic population $n_{(0)}(t)$ and its first order dynamical correction $n_{(1)}(t)$.
The adiabatic molecular population is obtained from the solution of the following  equation 
\begin{equation}\label{eq:occupation_0}
n_{(0)}(t) = \operatorname{Im} \left[ \int^{\infty}_{-\infty} \frac{d\omega}{2\pi} {{\mathcal {\widetilde G}}}^<_{(0)} (t,\omega) \right ].
\end{equation}
This equation is nonlinear since the expression for         {the} lesser Green's function under the integral depends on adiabatic electronic population $n_{(0)}(t)$; it is solved numerically by the bisection method.
It is known that depending on the parameters of the system, there can be several solutions to the zeroth order correction to the occupation \cite{galperin05}. We restrict ourselves to the situations with a single solutions.
The first order dynamical correction to the electronic occupation is obtained from {a} similar equation
\begin{equation}\label{eq:occupation_1}
n_{(1)}(t) = \operatorname{Im} \left[ \int^{\infty}_{-\infty} \frac{d\omega}{2\pi} {{\mathcal {\widetilde G}}}^<_{(1)} (t,\omega) \right ],
\end{equation}
where  ${\mathcal G}^<_{(1)}$ depends on $n_{(1)}$ and on         {the} already         {computed} value of $n_{(0)}$. Eq. (\ref{eq:occupation_1}) is linear and can be resolved analytically;         {an} analytical expression for $n_{(1)}(t) $  is given in appendix B.

Having computed time-dependent electronic occupation number
\begin{equation}
n(t) = n_{(0)}(t) +n_{(1)}(t),
\end{equation}
we calculate the correlation adiabatic (zeroth order) advanced or retarded  self-energies
\begin{equation}
{\widetilde \Sigma}^{A/R}_{c (0)} (T,\omega) = \chi n_{(0)}(T),
\end{equation}
and the first order dynamical corrections
\begin{equation}
{\widetilde \Sigma}^{A/R}_{c(1)} (T,\omega) = \chi n_{(1)}(T).
\end{equation}
Next,  we calculate Green's functions, adiabatic electric current  $J_{(0)}(t)$ using Eq.(\ref{eq:current_zeroth}) and first order dynamical correction $J_{(1)}(t)$ using Eq.(\ref{eq:current_first}).
Figure \ref{fig:plotMF5} shows         {the} adiabatic, zeroth order current; the current instantaneously follows the applied voltage and, as expected, does not demonstrate any dynamical features associated {     with non-adiabatic transport}. The polaron shift of the molecular orbital away from the voltage window, which  depends quadratically on strength or electron-vibration interaction $\lambda$, leads to the  current blockade around zero voltage. 

The time-dependent current computed with first order dynamical corrections is shown in Figure \ref{fig:plotMF6}. The current {     seemingly} shows a negative differential resistance behavior, {     with increasing current spikes as the voltage bias decreases and diminishing current at times of increasing voltage. These are observed to occur around when the oscillating right and left chemical potentials become equal.} The effect becomes  more pronounced  as         {the strength of the electron-vibration interaction $\lambda$ is increased.}

\begin{figure}
\includegraphics[width=1.0\columnwidth]{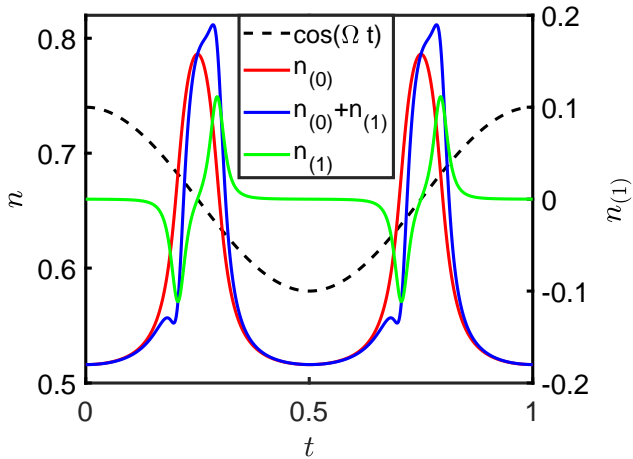}
\caption{Time-dependence of the adiabatic electronic occupation, electronic occupation with first order dynamical correction, and first order correction. Here $\chi = -0.8$, and the other parameters are the same as in Fig.  (\ref{fig:plotMF5}).}
    \label{fig:plotMF3}
\end{figure}

\begin{figure}
\includegraphics[width=1.0\columnwidth]{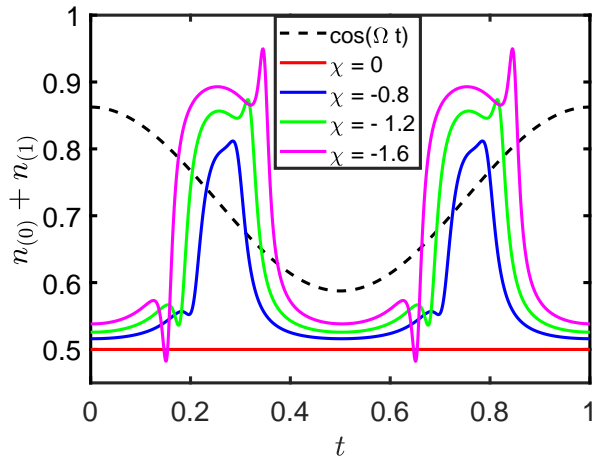}
\caption{Time-dependence of  electronic occupation computed with first order dynamical correction for various values of $\chi=- \frac{2\lambda^2}{\omega} $. The other parameters are the same as in Fig.  (\ref{fig:plotMF5}).}
\label{fig:plotMF2}
\end{figure}

To understand         {the} behavior of the time-dependent current, we turn our attention to         {the} time-evolution of         {the} electronic population         {in the} molecular junction. Figure \ref{fig:plotMF3}  shows adiabatic electronic populations $n_{(0)} (t)$, total electronic population $n(t) = n_{(0)}(t) +n_{(1)}(t) $  and first order corrections to electronic population  $n_{(1)} (t)$. From Figure \ref{fig:plotMF3}, we see that, just before the         {chemical potentials of the leads} meet, there is a dip within the electronic population, followed by a peak of equal magnitude whilst the lead's chemical         {potentials} diverge. This dynamical effect becomes more noticeable if the strength of         {the} electron-vibration coupling is         {increased,} as it is demonstrated in Figure \ref{fig:plotMF2}.

\begin{figure}
\includegraphics[width=1.0\columnwidth,scale=0.8]{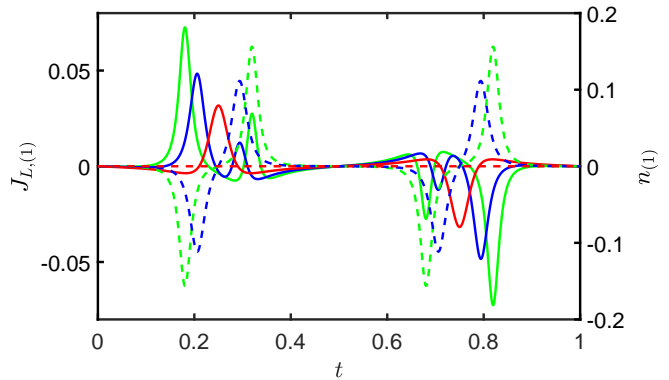}
\caption{Time-dependence of the first order dynamical corrections to the electronic population (dotted line) and to the current (full line). Here, the red lines correspond to $\chi = 0$; the blue to $\chi = -0.8$; and the green to $\chi = -1.2$. The other parameters are the same as in Fig.  (\ref{fig:plotMF5}). }
    \label{fig:plotMF8}
\end{figure}

Within the approximation employed to compute the correlation self-energy, the electronic energy level is shifted downward by the electron-vibrational coupling
\begin{equation}
\label{epst}
\epsilon(t) = \epsilon - \frac{2\lambda^2}{\omega}  ( n_{(0)}(t) +n_{(1)}(t)).
\end{equation}
In figure  \ref{fig:plotMF8} we plot side by side the first order dynamical corrections to electronic occupation numbers  $n_{(1)}(t)$ and electric current $J_{(1)}(t)$. One can see that the picks and dips in dynamical corrections to the electronic populations are mirrored by corresponding dips and picks in the time-dependence of the first order corrections to the current.
This  behavior can be easily interpreted  with the use of         {Eq.(\ref{epst}),} noticing that the dip in the electronic population temporarily brings the energy level into the voltage bias window, and therefore results         {in an increase in current}. Conversely, the peak in the electronic population pushes the electronic level {     down, out of resonance, thus resulting in the reduction in current.}

\section{Conclusion}

In this paper, we have developed a time-dependent transport theory for molecular junctions. The theory uses nonequilibrium Green's functions to separate time-scales on slow driving and fast electron tunneling. The gradient expansion of the Wigner space Kadanoff-Baym equations in terms of the slow time variable enabled us to convert the nonlocal in time integro-differential equations into a set of algebraic equations at each order of the gradient expansion. We found the solution of these         {equations}, and consequently produced analytic expressions for the first order dynamical corrections to the adiabatic Green's functions. The dynamical corrections to time-dependent electric current is obtained as well. The derived expressions are applicable to the general case of an AC electric current through a correlated central region. We applied the theory to two transport scenarios: a single-resonant level coupled to leads with harmonically oscillating chemical potentials; and         {a periodically} driven molecular junction coupled to a         {vibrational} mode described by the Holstein model.

For the noninteracting case, we found that the model agrees well with the exact results, within the realm of applicability. Upon the model's application to the Holstein model using time-dependent Hartree-approximation to treat electron-vibration         {interactions}, we found that the variation of the         {junction's} electronic population in time, due to nonadiabatic effects, induced small peaks  and dips within the current. This suggests that dynamical, nonlinear effects may have a significant influence on the current, depending on the         {central region's} correlations and the operating frequency of the junctions.    

Given the approximate nature of our approach, the conservation of charge deserves special attention.
Any meaningful theoretical approach should preserve,  at         each moment in time,  the continuity equation
\begin{equation} \label{eq:contEqn}
\frac{dN(t)}{dt} = J_R (t)  + J_L (t),
\end{equation}
 where 
\begin{equation} \label{eq:N(t)def}
{    
N(t) =  \operatorname{Im} \left[ \operatorname{Tr}\left[ \int^{\infty}_{-\infty} \frac{d\omega}{2\pi}  \bm{\mathcal {\widetilde G}}^<(t,\omega)  \right] \right]}
\end{equation}
is the occupation of the central region.
We have that $\bm{ \mathcal G}^< \approx \bm{ \mathcal{G}}^<_{(0)} + \bm{ \mathcal G}^<_{(1)}$. In the continuity equations, we disregard $\frac{d \bm{ \mathcal{G}}^<_{(1)}}{dt} \sim \bm{\mathcal G}^<_{(2)}$ and take into account that the adiabatic current is conserved $J_{R,(0)} (t)  + J_{L,(0)} (t) = 0$ , leaving the following continuity equation for the first order  dynamical corrections to the current:
\begin{equation}
\label{continuity}
{    
J_{R,(1)} (t)  + J_{L,(1)} (t)= \operatorname{Im} \left[ \operatorname{Tr}\left[ \int^{\infty}_{-\infty} \frac{d\omega}{2\pi} \frac{ d  \bm{ \mathcal G}^<_{(0)}}{dt} \right] \right].}
\end{equation}
In all considered cases we checked numerically that continuity equation (\ref{continuity}) is satisfied.

\begin{acknowledgements}
The authors would like to thank Vincent Kershaw and Riley Preston for many valuable discussions.
This work was supported by an Australian Government Research Training Program Scholarship to T.D.H.
\end{acknowledgements}

\clearpage

\appendix

\section{Self-energy for leads with AC voltage}
We assume  the  harmonic modulation of the leads' single-particle energies
\begin{equation}\label{eq:harm_kalpha}
\begin{split}
\epsilon_{k\alpha}(t) = \epsilon_{k\alpha} + \Delta_\alpha \cos\left(\Omega_\alpha t + \phi_\alpha \right),
\end{split}
\end{equation}
The self-energies of leads are
\begin{equation}
\label{eq:lead_self-energy}
\Sigma_{\alpha ij}^{R,A,<}(t,t') = \sum_{k} t^*_{k\alpha  i}(t) g_{k \alpha}^{R,A,< } (t,t')    t_{k \alpha j }(t'),
\end{equation}
where the Green's functions for separated leads are 

\begin{multline} 
g_{k \alpha}^{<}(t,t')= i f_{k \alpha}
\\
\times
e^{-i\epsilon_{k\alpha}(t-t')}e^{\frac{-2i\Delta_{\alpha}}{\Omega_\alpha}
\cos(\Omega_\alpha (t+t')/2 + \phi_\alpha)\sin(\Omega_\alpha (t-t')/2)}
\end{multline}
\begin{multline} 
g_{k \alpha}^{A}(t,t')=i\theta(t'-t) e^{-i\epsilon_{k\alpha}(t-t')}
\\
\times
e^{\frac{-2i\Delta_{\alpha}}{\Omega_\alpha}\cos(\Omega_\alpha (t+t')/2 + \phi_\alpha)\sin(\Omega_\alpha (t-t')/2)} , 
\end{multline}
\begin{equation} 
 g_{k \alpha}^{R}(t,t') =  g_{k \alpha}^{A}(t',t)^*.
\end{equation}


Introducing relative
$
\tau= t-t'
$
and central
$
T= (t+t')/2
$
times and 
using  the Jacobi-Anger expansion,
\begin{equation}
e^{iz\cos(\theta)}=\sum_{n=-\infty}^{\infty}i^{n}J_{n}(z)e^{in\theta},
\end{equation}
we bring free leads Green's function to the form
\begin{multline}
g_{k \alpha}^{A/R/<}(T,\tau) =  g_{k \alpha}^{A/R/<}(\tau) \\
\times \sum_{n=-\infty}^{\infty}(-1)^{n}J_{n} \left(\psi_\alpha(T) \right) e^{\frac{i}{2}n\Omega_\alpha \tau},
\end{multline}
where
\begin{equation}
\psi_\alpha(T) =\frac{2\Delta_{\alpha}}{\Omega_\alpha}\cos(\Omega_\alpha T + \phi_\alpha) 
\end{equation}
and  $g_{k \alpha}^{A/R/<}(\tau)$ are standard advanced, retarded, and lesser leads Green's function which depend on relative time only and are computed for static leads:
\begin{equation}
g_{k \alpha}^{<}(\tau)= i f_{k \alpha}e^{-i\epsilon_{k\alpha}(\tau)}, 
\end{equation}
\begin{equation}
g_{k \alpha}^{A}(\tau)=i\theta(-\tau) e^{-i\epsilon_{k\alpha}(\tau)},
\end{equation}
\begin{equation}
g_{k \alpha}^{R}(\tau) =  g_{k \alpha}^{A}(-\tau)^*.
\end{equation}

The substitution of these Green's functions   into the definition of lead self-energies (Eq. (\ref{eq:lead_self-energy})) gives us 
\begin{multline}
\mathbf \Sigma_{\alpha}^{A/R/<}(T,\tau)
=\mathbf \Sigma_{\alpha}^{A/R/<}(\tau) 
\\
\times \sum_{n=-\infty}^{\infty}(-1)^{n}J_{n}\left( \psi_\alpha(T)\right) e^{\frac{i}{2}n\Omega_\alpha \tau}
\end{multline}
where  $\mathbf \Sigma_{\alpha}^{A/R/<}(\tau)$ is again the standard self-energy matrix computed for static leads. For the solutions of {the Kadanoff-Baym equations} we will need self-energies computed in Wigner space
\begin{equation}
\widetilde{\mathbf \Sigma}_{\alpha}^{A/R/<}(T,\omega)=\int_{-\infty}^{\infty}d\tau e^{i\omega\tau} \mathbf \Sigma_{\alpha}^{A/R/<}(T,\tau).
\end{equation}
The direct {calculation} gives the lesser self-energy
\begin{multline}\label{eq:defn less self energy for wigner space}
\widetilde{\mathbf \Sigma}_{\alpha}^{<}(T,\omega)=i\sum_{n=-\infty}^{\infty}(-1)^{n}f_{\alpha}(\omega+n\Omega_\alpha /2) \mathbf \Gamma_{\alpha}(\omega+n\Omega_\alpha/2) 
\\
\times J_{n}\left(\psi_\alpha(T)\right).
\end{multline}
and advanced/retarded lead self-energies ($\pm$ for the advanced (retarded) self-energies):
\begin{multline}
\label{eq: defn A/R self-energy for wigner space}
\widetilde{\mathbf \Sigma}_{\alpha}^{A/R}(T,\omega)=\sum_{n=-\infty}^{\infty}(-1)^{n} \Big[\mathbf \Lambda_\alpha \left(\omega+\frac{n}{2}\Omega_\alpha \right) 
\\
\pm \frac{i}{2} \mathbf \Gamma_\alpha \left(\omega+\frac{n}{2}\Omega_\alpha \right) \Big] J_{n}\left(\psi_\alpha(T)\right).
\end{multline}
The  level-width functions are
\begin{equation}
 \Gamma_{\alpha ij } (\omega) = 2 \pi \sum_{k} \delta(\omega-\epsilon_{k\alpha})  t^*_{k\alpha i} t_{k\alpha j}.
 \end{equation}
 
The level-shift functions $\Lambda_{\alpha ij }(\omega)$ can be computed from $\Gamma_{\alpha ij}(\omega)$  via Kramers-Kronig relation. To simplify analysis, we make the wide-band approximation, assuming that the leads density of states is constant, that means that the level-width function is a constant, $\omega$ independent matrix
\begin{equation}
\mathbf  \Gamma_{\alpha } (\omega)  = \mathbf  \Gamma_{\alpha },
 \end{equation}
and that  the level-shift functions         {vanishes}
\begin{equation}
\mathbf  \Lambda_{\alpha } (\omega)  = 0.
 \end{equation}

 This gives us the following self-energies:
\begin{multline}
\widetilde{\mathbf \Sigma}_{\alpha}^{<}(T,\omega)=i \mathbf \Gamma_{\alpha}  \sum (-)^{n}f_{\alpha}(\omega+n\Omega_\alpha /2)
 J_{n}\left(\psi_\alpha(T)\right),
\end{multline}

\begin{equation}
\widetilde{\mathbf \Sigma}_{\alpha}^{A}(T,\omega)=\frac{i}{2} \mathbf \Gamma_{\alpha},
\end{equation}

\begin{equation}
\widetilde{\mathbf \Sigma}_{\alpha}^{R}(T,\omega)=-\frac{i}{2} \mathbf \Gamma_{\alpha}.
\end{equation}

The time and energy derivatives of the lesser leads self-energy in Wigner space, which are required for the gradient expansion of Kadanoff-Baym equation, are
\begin{multline}
\partial_T \bm \Sigma^<_\alpha = i\bm\Gamma_{\alpha} \Delta_\alpha \sin \left(\Omega_\alpha T \right)  \sum_{n=-\infty}^{\infty}(-)^{n}f_{\alpha}(\omega+n\Omega_\alpha /2) \\
\times \Big[ J_{n+1}\left(\psi_\alpha(T)\right) 
- J_{n-1}\left(\psi_\alpha(T)\right) \Big],
\end{multline}

\begin{multline}
\partial_\omega \bm \Sigma^<_\alpha = -i{\displaystyle \frac{\bm \Gamma_{\alpha}}{\zeta_\alpha}\sum_{n=-\infty}^{\infty}}(-)^{n}\left[ f_{\alpha}\left(\omega+\frac{1}{2}n\Omega_\alpha \right) \right]^{2}
\\
\times e^{\left(\omega+\frac{1}{2}n\Omega_\alpha-\mu_\alpha\right)/\zeta_\alpha}J_{n}\left(\psi_\alpha(T)\right).
\end{multline}

\section{First order dynamical corrections to electronic population in Holstein model}

The first correction to the occupation can be related to the adiabatic result. Making use of Eqs. (\ref{eq:occupation_corrections}) and (\ref{eq:G^<_{(1)}}), we find
\begin{multline} \label{eq:occupation_firstMF}
n_{(1)}(T) =  \int^{\infty}_{-\infty} \frac{d\omega}{2\pi} \operatorname{Im} \Big[  -\frac{i}{2} \widetilde{{\mathcal G}}^{R}_{(0)} \partial_T \widetilde{{\mathcal G}}^{<}_{(0)}  
\\
-\frac{i}{2} \widetilde{{\mathcal G}}^{R}_{(0)} \partial_T  {\widetilde{\Sigma}}^{<} \partial_\omega\widetilde{{\mathcal G}}^{A}_{(0)}
+\frac{i}{2} \widetilde{{\mathcal G}}^{R}_{(0)} \partial_\omega {\widetilde{\Sigma}}^{<} \partial_T \widetilde{{\mathcal G}}^{A}_{(0)}
\\
-\frac{i}{2} \widetilde{{\mathcal G}}^{R}_{(0)} \chi \partial_{T} n_{(0)} \partial_{\omega}\widetilde{{\mathcal G}}^{<}_{(0)}
+  \widetilde{{\mathcal G}}^{R}_{(0)} {\widetilde{\Sigma}}^{<} \widetilde{{\mathcal G}}^{A}_{(1)} \Big].
\end{multline} 
Here, the first correction to the advanced Green's function is given as 
\begin{equation} \label{eq:GA/R_firstMF}
\begin{split}
\widetilde{{\mathcal G}}^{A/R}_{(1)} 
=\widetilde{{\mathcal G}}^{A/R}_{(0)} \chi n_{(1)} \widetilde{{\mathcal G}}^{A/R}_{(0)}. 
\end{split}
\end{equation}
For Eq. (\ref{eq:occupation_firstMF}), we observe that the result can be split into contributions that depend on the adiabatic solution only and the term containing $n_{(1)}$ itself.  Hence making use of the Eqs. (\ref{eq:occupation_firstMF}) and (\ref{eq:GA/R_firstMF}), we find a linear equation such that: 
\begin{equation}
\begin{split}
n_{(1)}(T) =  \frac{A(T)}{1-B(T)},
\end{split}
\end{equation}
where

\begin{equation}
\begin{split}
A(T) =  \int^{\infty}_{-\infty} \frac{d\omega}{2\pi}\operatorname{Im} \left [  -\frac{i}{2} \widetilde{{\mathcal G}}^{R}_{(0)} \partial_T \widetilde{{\mathcal G}}^{<}_{(0)} 
-\frac{i}{2} \widetilde{{\mathcal G}}^{R}_{(0)} \partial_T  {\widetilde{\Sigma}}^{<} \partial_\omega\widetilde{{\mathcal G}}^{A}_{(0)}
\right.
\\
\left.
+\frac{i}{2} \widetilde{{\mathcal G}}^{R}_{(0)} \partial_\omega {\widetilde{\Sigma}}^{<} \partial_T \widetilde{{\mathcal G}}^{A}_{(0)}
-\frac{i}{2} \widetilde{{\mathcal G}}^{R}_{(0)} \chi \partial_{T} n_{(0)} \partial_{\omega}\widetilde{{\mathcal G}}^{<}_{(0)}\right],
\end{split}
\end{equation}

\begin{equation}
\begin{split}
B(T) = \chi  \int^{\infty}_{-\infty} \frac{d\omega}{2\pi}\operatorname{Im} \left [  \widetilde{{\mathcal G}}^{R}_{(0)} {\widetilde{\Sigma}}^{<} \left(\widetilde{{\mathcal G}}^{A}_{(0)}\right)^2 \right].
\end{split}
\end{equation}

\clearpage


%

\end{document}